# Optimal FX Hedge Tenor with Liquidity Risk

Rongju Zhang

Postdoctoral Fellow, Risklab, Data61, CSIRO, AU and Adjunct Research Fellow, Centre for Quantitative Finance and Investment Strategies, Monash University, AU

Email: henry.zhang@data61.csiro.au

Mark Aarons

Adjunct Associate Professor, Centre for Quantitative Finance and Investment Strategies, Monash University, AU and Head of Investment Risk, VFMC, AU

Email: maarons@me.com

Gregoire Loeper

School of Mathematical Sciences and Centre for Quantitative Finance and Investment Strategies, Monash University, AU

Email: gregoire.loeper@monash.edu

March 14, 2019

*Acknowledgements: the authors wish to thank IFM Investors for their practical perspective and for supplying useful data for this research. The Centre for Quantitative Finance and Investment Strategies is supported by BNP Paribas. The authors also wish to thank Michael Landman for his valuable comments and remarks.*

## Abstract

We develop an optimal currency hedging strategy for fund managers who own foreign assets to choose the hedge tenors that maximize their FX carry returns within a liquidity risk constraint. The strategy assumes that the offshore assets are fully hedged with FX forwards. The chosen liquidity risk metric is Cash Flow at Risk (CFaR). The strategy involves time-dispersing the total nominal hedge value into future time buckets to maximize (minimize) the expected FX carry benefit (cost), given the constraint that the CFaRs in all the future time buckets do not breach a predetermined liquidity budget. We demonstrate the methodology via an illustrative example where shorter-dated forwards are assumed to deliver higher carry trade returns (motivated by the historical experience where AUD is the domestic currency and USD is the foreign currency). We also introduce a tenor-ranking method which is useful when this assumption fails. We show by Monte Carlo simulation and by backtesting that our hedging strategy successfully operates within the liquidity budget. We provide practical insights on when and why fund managers should choose short-dated or long-dated tenors.

## 1. Introduction

Optimising investment returns subject to risk constraints is one of the key functions of every fund manager. One risk constraint which is of great importance, but has received scant attention in the literature, is liquidity risk generated by currency hedging. Currency hedging via FX forward contracts generates cash outflows when the domestic currency falls relative to the currency of the offshore assets being hedged. If the FX forwards are uncollateralised, which is commonly the case, those outflows will occur at the settlement date of the contract. At that time, the fund manager must ensure that they have sufficient cash to fund the settlement amount. This can be achieved by various means including: (a) holding cash; (b) selling assets prior to the settlement date to generate cash; (c) borrowing cash; or (d) repoing bonds[1] in exchange for cash. All options have costs. Option (a) generates a cash drag, whilst option (b) involves the risk of having to sell liquid assets when their prices are low. This is especially the case if the domestic currency is positively correlated to the risky asset such that both drop in value at the same time, which is typically the case for AUD/USD for example. Option (b) becomes more difficult if the fund manager holds a high proportion of illiquid assets. In such a case, there may not be sufficient liquid assets to sell if the domestic currency falls by a large enough amount. Option (c) may not be permitted in many cases as leverage can be outside the mandate of many funds. And option (d) also has a cost and only works to the extent that eligible bonds are available in sufficient quantity.

There are two ways in which this liquidity risk can be mitigated. The most obvious way is to reduce the FX-hedge ratio. Whilst this may be suitable for some fund managers at certain times, other fund managers may not have this choice (e.g. due to their mandates) or may not wish to make this choice for numerous reasons[2]. The second way to mitigate this liquidity risk is by transacting many FX forwards with smaller principals and staggering their settlement dates over a sufficiently long period of time. This ensures that, in a given time window, only a small number of FX forwards will settle, proportionately reducing the cash required to fund their settlement. For instance, instead of rolling over a single FX forward with a principal of USD $12m expiring in every three month time, a fund could transact twelve FX forwards with a principal of USD $1m with expiry dates staggered quarterly over three years, thus reducing the liquidity risk in each quarterly period by 11/12.

However, transacting longer-dated uncollateralised FX forwards involves a trade-off as they have higher transaction costs. This is because the banks, who sell FX forwards to funds, must charge for credit value adjustment (CVA), the cost of capital and bid/offer spread, all of which increase monotonically with tenor. A further and significant tenor-related trade-off is that FX forwards generate a carry cost or benefit which is tenor-dependent.

The question therefore arises: what is the optimum tenor over which FX forwards should be staggered to minimise transaction costs, maximise (minimize) any carry benefit (cost) but remain within a given liquidity risk budget? Answering this question is of considerable practical relevance. Funds which transact excessively long FX forwards might be paying unnecessarily high transaction costs and foregoing FX carry benefit, whilst funds which transact excessively short FX forwards risk taking on excessive liquidity risk.

The latter case occurred to the MTAA superannuation fund in Australia in late 2008[3]. In that case the fund, which was 45% invested in illiquid assets, was hedging its offshore assets using very short-dated FX forwards. When the Australian dollar plunged over 30% against the US dollar in the space of several months, it had to fund large hedge settlement obligations. To generate the required cash, it sold liquid assets at the worst possible time – since equities had also declined very substantially in value. As it sold liquid assets, MTAA's proportion of illiquid assets soared to 70%. At that point, the fund decided to cease currency hedging in case the AUD kept falling. Unfortunately for MTAA and its members, the AUD rebounded shortly the thereafter, creating massive losses as the now unhedged offshore assets declined sharply in value in AUD terms.

The importance of managing currency hedging well is increasing as fund managers globally are allocating a larger proportion of their funds to offshore assets. In addition, an increasing proportion of these funds are being allocated to illiquid investments in the form of infrastructure, property, hedge funds, private credit and private equity. Thus having a systematic framework for currency hedging is of considerable practical importance for fund managers.

---

[1] A repo or repurchase agreement is where bonds or other assets are sold in exchange for cash, then repurchased at a later date.
[2] For example, an infrastructure fund manager whose skill is purely in selecting and managing global infrastructure assets may want to remove all FX exposure.
[3] The source is available in https://www.afr.com/business/banking-and-finance/mtaa-supers-comeback-from-liquidity-troubles-in-2008-to-top-super-fund-in-2015-20160120-gma2zb.

In this paper we propose and implement a rigorous methodology for determining the optimum tenor over which FX forwards should be staggered to ensure that a liquidity constraint can be satisfied to a specified statistical confidence level, whilst minimizing the transaction costs and maximizing any carry benefit. The liquidity constraint is specified in terms of cash flow at risk (CFaR) which is a simple and relatively common liquidity risk metric used by corporate treasury departments[4].

This methodology is built upon a long-run mean-reversion assumption on the FX spot rate, from which the expressions of expected carry and CFaR can be analytically derived. To find the optimal combinations of hedge tenors, the first step is to evaluate all the CFaRs in the future time buckets to identify if the CFaRs have breached the liquidity budget. The second step is to fill up the CFaR capacities, starting from the shortest tenor and incrementally increasing it to the longest tenor, or starting from the tenor with the highest (expected) carry to the tenor with the lowest (expected) carry, by allocating the nominal hedge values to the tenors in which the CFaRs are under the liquidity budget, until the total hedge value has been fully allocated. The resulting optimal tenors will be the ones that maximize the expected carry return with possible transaction costs subject to the liquidity constraint.

## 2. Key parameters

Although our methodology is general, for illustrative purposes we assume that AUD is the home currency and USD is the foreign currency.

### Liquidity and Cash Flow at Risk

Fund managers usually need to hold a cash account to prepare for any potential cash outflows. If the cash reserve is not sufficient for offsetting a cash outflow, they may be required to liquidate their assets at undesirable prices, whilst if the cash reserve is unnecessarily high, they may suffer reduced investment returns.

The cash reserve can be interpreted as liquidity budget. We aim to ensure that the cash reserve is sufficient for covering a tail event of the cash outflow which is defined by CFaR. The cash flows generated from the currency hedging can be accumulated in the cash reserve. In practice, investment funds may pay out part of the cash reserve to investors if the cash reserve level is sufficiently high (due to cash inflows) and tend to top up the cash reserve if it is too low (due to cash outflows), so as to maintain the cash reserve level within a certain range, usually between 1% and 10% proportional to the fund size. We assume, for simplicity, that this proportional cash reserve level (the liquidity budget) is constant over time.

### Carry and tenor dependence

Currency carry trades are very simple: one sells a lower yielding currency and buys a higher yielding one to earn the interest rate differential. Traders typically execute this via FX forward contracts. Currency carry trades can persistently generate positive excess returns over long periods, but they can also suffer severe losses in short periods of time. The key risk is that the domestic currency crashes during the trade, more than offsetting any gains from the interest rate differential and cross-currency basis (the so-called peso problem). Currency carry trades have received a great deal of attention in the academic literature over the last three decades. The original work on carry trade in the literature traces back to Hansen and Hodrick (1980) who reject the hypothesis that the expected return to speculation in the FX forward market is zero. Fama (1984) finds time-varying premium in forward rates, and the premium and expected change in the spot rate are negatively correlated. Brunnermeier, Nagel and Pedersen (2008) show by analyzing eight currencies (including AUD) against USD, that the carry trade returns are negatively skewed and carry traders are subject to crash risk. Their findings also motivate the need for managing liquidity risk for the carry trade. In recent years, numerous aspects have been studied, including why the violation of uncovered interest rate parity (UIP) exists (see for example Jurek (2014)), how positive carry can be predicted (see for example Chen (2017)) and how the currency carry trade can be enhanced (see for example Berge, Jorda and Taylor (2010)).

---

[4] See https://www.risk.net/definition/cashflow-risk

In this paper, we take a different approach to currency carry trade. The conventional literature deals with participants for whom the carry trade is a voluntary, speculative activity. Instead, we will focus on fund managers who are required, or choose, to FX-hedge 100% of their offshore assets. From an Australian perspective, where until recently AUD interest rates were persistently higher than USD interest rates, this currency hedging is effectively an involuntary currency carry trade.

Furthermore, we decompose the currency carry trade into two independent components: (a) the spot FX component; and (b) the interest rate and cross-currency basis[5] component. For fund managers who are 100% FX-hedged, the spot FX component generates liquidity risk, but does not otherwise contribute to any gains or losses in overall fund NAV since any fall in the spot FX rate is exactly offset by a corresponding rise in the value of the offshore assets being hedged. In contrast, the interest rate and basis component does generate gains or losses in the net asset value (NAV) of the fund. The question then becomes: how to maximise the gains or minimise the losses of this component? The key factor in this regard is the *tenor* over which the hedging occurs. Fund managers who are 100% FX-hedged cannot choose the timing of their carry trades nor the currency pair(s) involved. They can, however, choose whether to hedge using short, medium or long-dated FX forwards. Since the interest rate differential and cross-currency basis are both tenor-dependent, this choice matters. Figure 1 shows that in the Australian context, hedging using shorter-dated FX forwards between 1993 and 2018 would have generated higher returns than using longer-dated FX forwards. Figure 2 and Figure 3 further decompose the contribution to the carry trade return into the spot component and the FX forward point component. The figures show that over long periods the FX forward point contribution is dominant although that over short/medium periods the noise from spot FX can swamp the slow, steady upward drift from earning the FX forward points. During this 25 year period, the carry trade return is dominated by the FX forward points, and the forward curve was consistently steeper at the short-end and become flatter further out. As a result, rolling short-dated hedges was, with the benefit of hindsight, preferably to a long-dated hedge, notwithstanding the continual movements of the forward curve. Therefore, in this paper, we assume that FX hedging with shorter-dated tenors tend to deliver high returns.

The tenor dependence of the currency carry trade does not appear to have been previously addressed in the academic literature. All previous papers have focussed on short-dated – typically one month – rolling carry trades. In our framework, the tenor dependence is critical. Not only can it have a structural impact on the quantum of returns from currency hedging, but it is also a key contributor to liquidity risk. For a fixed value of nominal hedge, short-dated hedges generate a much higher CFaR than hedges whose maturity dates are staggered over longer periods.

Whilst the framework presented in this paper is entirely general, our examples are inspired by the Australian context where rolling shorter-dated hedges versus USD generate higher returns. Shorter-dated hedges also have lower transaction costs. However, this poses a key challenge which is the main subject of this paper. Given that hedging short is cheaper, whilst hedging long reduces liquidity risk, what is the "optimum" tenor for currency hedging? Our framework provides a rigorous approach to answering this question. In the Australian context, it allows one to compute the shortest and hence cheapest tenor for FX hedging whilst still maintaining a specified level of CFaR.

---

[5] Cross-currency basis is the market price for term-hedging currency risk. It is tenor dependent and currency pair dependent, and may be positive or negative.

**Figure 1** Cumulative Carry Trade P&L from Rolling Hedging 1 unit of USD – Total Return

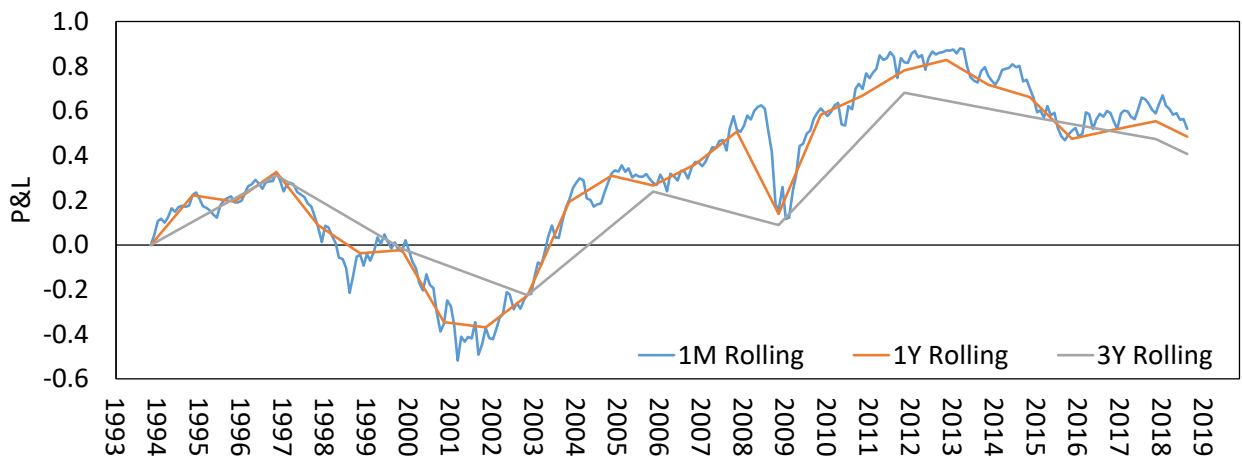

**Figure 2** Cumulative Carry Trade P&L from Rolling Hedging 1 unit of USD – Spot Component

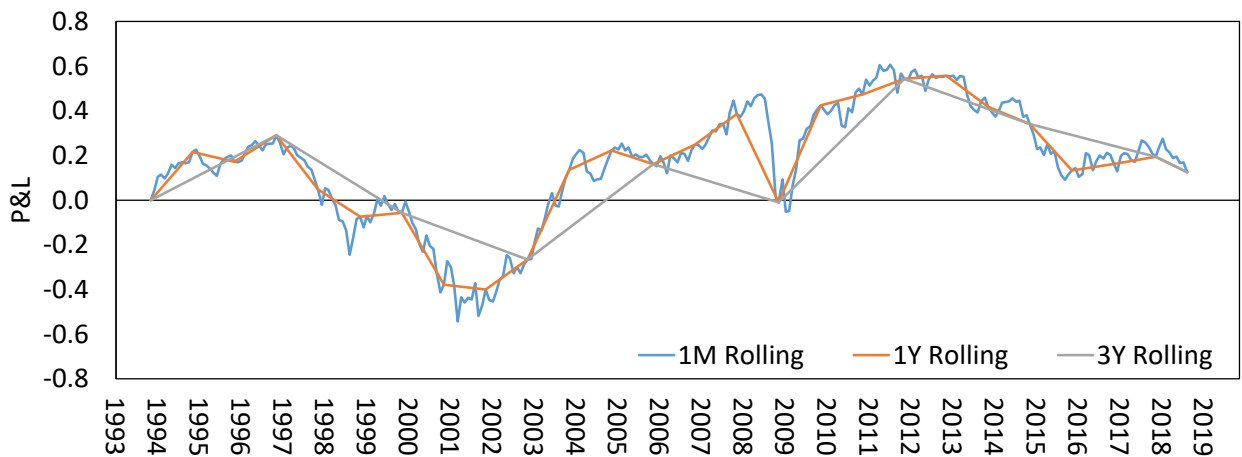

**Figure 3** Cumulative Carry Trade P&L from Rolling Hedging 1 unit of USD – Forward Points Component

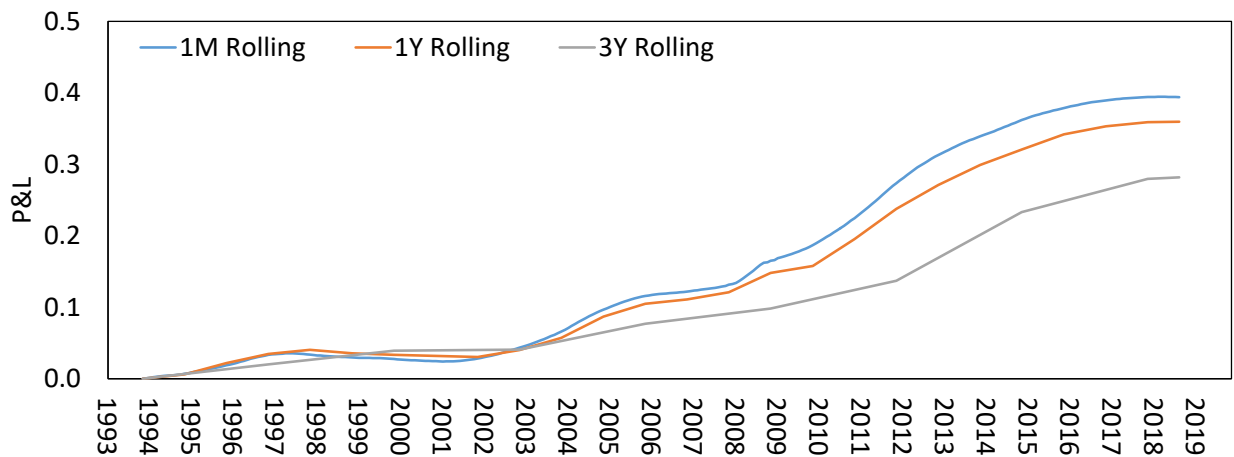

# 3. The Problem

Let $W_t$ be the asset value in foreign currency. Denote respectively by $W_t^U$ and $W_t^H$ the unhedged and hedged value in the offshore asset. So, we have $W_t = W_t^U + W_t^H$ and we assume $W_0^U = W_0$ and $W_0^H = 0$. In this paper, we aim to maintain the fully hedged position for the asset value $W_t$, and new FX forwards are transacted once old ones mature – whenever $W_t^U > 0$, an equivalent nominal hedge amount, denoted by $A_t$, will be entered to offset the value of $W_t^U$. At the initial time, the value to be hedged is simply $A_0 = W_0^U = W_0$. Denote by $a_{t,T}$ the nominal amount of the hedge in foreign currency bought at time $t$ expiring at time $T$. We do not hedge the value in one single tenor; instead, we spread out the total nominal hedge into different hedge tenors, so the hedges entered at time $t$ with different expiries at times $T > t$ are $(a_{t,T})_{T>t}$. At any time $t > 0$, the forward contracts with nominal amounts $(a_{\tau,t})_{\tau<t}$ will expire, so there will be $W_t^U = \sum_{\tau<t} a_{\tau,t}$ in the asset value that becomes unhedged. To maintain the fully hedged requirement, we need to hedge the value of $A_t = \sum_{\tau<t} a_{\tau,t}$ by taking long positions in the domestic currency. In this paper, we assume but in general do not restrict that the asset value $W_t = W_0$ constant over time. Removing the variation in the asset value allows us to focus on the analysis and effects derived purely from the perspective of currency hedging.

Denote by $S_t$ the spot rate of the units of the foreign currency per unit of the domestic currency. Denote by $F_{t,T}$ the forward rate at time $t$ expiring at time $T$, so the tenor of such forward contract is $T - t$. The payoff at time $T$ of the forward contract transacted at time $t$ in domestic currency at maturity is $F_{t,T} - S_T$. The annualized carry trade return from this forward contract at its expiry will be $(F_{t,T} - S_T)/(T-t)$ in domestic currency. The cash flow in domestic currency at time $t$, denoted by $\text{CF}_t$, will be the sum of all the cash flows for the forward contracts that were purchased previously in different times with the same expiry at time $t$, i.e.,

$$\text{CF}_t = \sum_{\tau<t} a_{\tau,t}\left(F_{\tau,t} - S_t\right). \tag{1}$$

We also aim to satisfy some liquidity criteria of the hedging program. We define the liquidity risk metric as the Value at Risk (VaR) of a cash flow, also known as the Cash Flow at Risk (CFaR). Mathematically, the CFaR is defined as the $(1-p)$ quantile of the random variable $-\text{CF}$. Denote by $p\text{-CFaR}_{T|t}$ the $p$ quantile of CFaR at time $T$ conditional on the information available at time $t$. Denote by $L$, measured in domestic currency, the liquidity budget that is reserved for covering potential cash outflows. The purpose at time $t$ is to ensure that the liquidity budget $L$ is sufficient to cover a tail event of a cash outflow, $p\text{-CFaR}_{T|t}$, i.e.,

$$p\text{-CFaR}_{T|t} \leq L, \tag{2}$$

for all the future times $T = t+1, t+2, \ldots$ .

Finally, the optimal combination of hedge tenors is determined by maximizing the expected annualized carry trade return, subject to the fully hedged constraint and the liquidity constraint. The optimization objective is

$$\begin{aligned}
\max_{(a_{t,T})_{T>t}} \quad & \sum_{T>t} a_{t,T} \frac{F_{t,T} - E_t[S_T]}{T-t}, \\
\text{s.t.} \quad & p\text{-CFaR}_{T|t} \leq L, \text{ for } T > t, \\
& \sum_{T>t} a_{t,T} = A_t.
\end{aligned} \tag{3}$$

A similar optimization objective is used for portfolio optimization in Alexander and Baptista (2002). The main difference is that our approach manages the tail risk for all the future time steps, whilst the conventional mean-VaR approach only manages the risk one-step ahead.

Throughout this paper, we measure the forward payoff, cash flow, CFaR and $L$ in domestic currency, and we measure the asset value and nominal hedge amount in foreign currency.

## 4. Stochastic Modelling

Empirical research show that FX spot rates exhibit mean-reverting property, see for example Engel and Hamilton (1990) and Sweeney (2006). In this paper, we use a very simple one factor process, namely Ornstein Uhlenbeck process, to model the spot rate. The process is

$$dS_t = k(\theta - S_t)dt + v dB_t, \qquad (4)$$

where $\theta$ represents the long-run mean of the spot rate, $k$ represents the mean-reversion speed of the spot rate, $v$ represents the volatility of the spot rate, and $B_t$ is the Brownian motion.

The conditional mean of the spot rate is

$$E_t(S_T) = S_t e^{-k(T-t)} + \theta\left(1 - e^{-k(T-t)}\right), \qquad (5)$$

and the conditional variance is

$$\sigma_t(S_T)^2 = \text{var}_t(S_T) = \frac{v^2}{2\kappa}\left(1 - e^{-2k(T-t)}\right). \qquad (6)$$

The cash flow at time $T$ conditional on time $t$ follows a normal distribution with expectation

$$\begin{aligned} E_t(\text{CF}_T) &= E_t\left(\sum_{\tau \leq t} a_{\tau,T}(F_{\tau,T} - S_T)\right) \\ &= \sum_{\tau \leq t} a_{\tau,T}(F_{\tau,T} - E_t(S_T)), \end{aligned} \qquad (7)$$

and variance

$$\begin{aligned} \sigma_t(\text{CF}_T)^2 &= \text{var}_t(\text{CF}_T) \\ &= \text{var}_t\left(\sum_{\tau \leq t} a_{\tau,T}(F_{\tau,T} - S_T)\right) \\ &= \left(\sum_{\tau \leq t} a_{\tau,T}\right)^2 \sigma_t(S_T)^2. \end{aligned} \qquad (8)$$

Note that these conditional moments are defined conditional on the spot price at time $t$ as well as the hedge contracts entered before and including time $t$. Using the properties of normal distribution, we obtain the expression of the CFaR at time $T$ conditional on time $t$ given by

$$\begin{aligned} p\text{-CFaR}_{T|t} &= -E_t(\text{CF}_T) - \sigma_t(\text{CF}_T)\Phi^{-1}(p) \\ &= -\sum_{\tau \leq t} a_{\tau,T}(F_{\tau,T} - E_t(S_T)) - \left(\sum_{\tau \leq t} a_{\tau,T}\right)\sigma_t(S_T)\Phi^{-1}(p). \end{aligned} \qquad (9)$$

From this equation, we can derive

$$p\text{-CFaR}_{T|t} = p\text{-CFaR}_{T|t^-} + a_{t,T} \times p\text{-CFaR}_{T|t}^{\text{Unit}}. \qquad (10)$$

Here,

$$p\text{-CFaR}_{T|t^-} = -\sum_{\tau \le t-1} a_{\tau,T}\left(F_{\tau,T} - E_t(S_T)\right) - \left(\sum_{\tau \le t-1} a_{\tau,T}\right)\sigma_t(S_T)\Phi^{-1}(p), \tag{11}$$

represents the CFaR at time $T$ conditional on the time immediately before the new hedge at time $t$ is entered. We denote by $t^-$ the time immediately before time $t$. The second term $a_{t,T} \times p\text{-CFaR}_{T|t}^{\text{Unit}}$ represents the incremental CFaR due to the new forward contracts entered at time $t$, where

$$p\text{-CFaR}_{T|t}^{\text{Unit}} = -\left(F_{t,T} - E_t(S_T)\right) - \sigma_t(S_T)\Phi^{-1}(p), \tag{12}$$

is the CFaR for hedging one unit of asset value in the foreign currency.

## 5. Optimization and the optimal hedge strategy

The optimization problem becomes

$$\begin{aligned}
\max_{(a_{t,T})_{T>t}} \quad & \sum_{T>t} a_{t,T} \frac{F_{t,T} - E_t[S_T]}{T-t}, \\
\text{s.t.} \quad & a_{t,T} \le \frac{L - p\text{-CFaR}_{T|t^-}}{p\text{-CFaR}_{T|t}^{\text{Unit}}}, \text{ for } T > t, \\
& \sum_{T>t} a_{t,T} = A_t.
\end{aligned} \tag{13}$$

At any time $t$, we want to determine an optimal combination of hedge tenors $(a_{t,T})_{T<t}$ to hedge a value of $A_t$ in foreign currency, given that $E_t[S_T]$, $p\text{-CFaR}_{T|t^-}$ and $p\text{-CFaR}_{T|t}^{\text{Unit}}$ can be calculated according to the Ornstein Uhlenbeck assumption on the spot process.

Our empirical findings show that, historically on average, for investors hedging USD assets back to AUD, that short-dated forwards tend to deliver higher returns from carry trade than hedging with long-dated forwards. Therefore, we assume that

$$\frac{F_{t,T} - E_t[S_T]}{T-t} > \frac{F_{t,\tau} - E_t[S_\tau]}{\tau-t} \text{ for } t < T < \tau. \tag{14}$$

This assumption can be relaxed and we will discuss more details later in this paper.

Given such assumption, we maximize the nominal hedge amounts for the shortest tenors, subject to the fully hedged constraint and the liquidity risk constraint.

There are four key steps to solve this optimization problem:
a. Calculate $p\text{-CFaR}_{T|t^-}$ for all $T > t$ according to the previously hedged amounts.
b. Identify if any of $p\text{-CFaR}_{T|t^-}$ for all $T > t$ have breached the liquidity budget $L$.
c. Use negative hedges (short domestic currency) to reduce the breached CFaRs.

   1. For those that have breached the liquidity budget $L$, i.e., $p\text{-CFaR}_{T|t^-} > L$ for any $T > t$, enter a negative hedge position $a_{t,T} < 0$ to reduce the overall CFaR to the liquidity budget $L$, i.e.,

   $$L = p\text{-CFaR}_{T|t^-} + a_{t,T} \times p\text{-CFaR}_{T|t}^{\text{Unit}}. \tag{15}$$

   Therefore, we have

   $$a_{t,T} = \frac{L - p\text{-CFaR}_{T|t^-}}{p\text{-CFaR}_{T|t}^{\text{Unit}}}. \tag{16}$$

2. Update the prevailing total hedge amount with these new negative hedges, such that the new nominal hedge amount will be

$$A_t + \sum_{T \in \{\tau : a_{t,\tau} < 0\}} |a_{t,T}|. \tag{17}$$

d. Hedge the amount $A_t + \sum_{T \in \{\tau : a_{t,\tau} < 0\}} |a_{t,T}|$.

1. Find the shortest tenor for which the liquidity constraint has not been breached

$$T^* = \min_{T > t} \{T : p\text{-CFaR}_{T|t^-} < L\}. \tag{18}$$

2. Hedge the amount for this tenor such that the resulting CFaR matches the liquidity budget, i.e.,

$$a_{t,T^*} = \frac{L - p\text{-CFaR}_{T^*|t^-}}{p\text{-CFaR}^{\text{Unit}}_{T^*|t}}. \tag{19}$$

3. If the fully hedged constraint has not been breached, repeat (i) and (ii), otherwise, adjust the nominal hedge $a_{t,T^*}$ such that the fully hedged constraint can be maintained.

## 6. An Illustrative Example

The objective is to use the shortest tenors and maintain the CFaRs at the liquidity budget $L=1$ which is set to be one unit of domestic currency in this illustrative example. Suppose now we want to enter forwards to remove currency exposure of a certain unhedged amount of asset value due to the expiry of the previous hedges. We want to hedge this amount to maintain a 100% hedge ratio. The four key steps in the algorithm are explained in the following.

Step 1: Calculate the CFaRs in all the future times based on the forward contracts that were purchased previously. Suppose in this example, at the current time, time 0, we only have the forward contracts expiring up to time step 8, so that the CFaRs at time step 9 and onwards are zero.

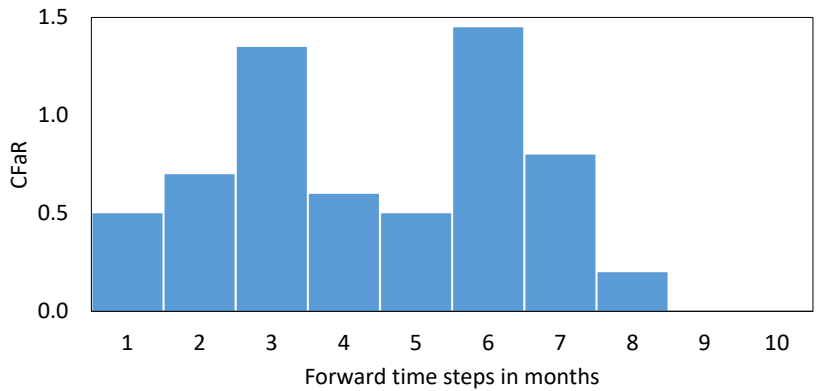

Step 2: Determine whether any of these CFaRs has breached the liquidity constraint. In this example, the CFaRs at time steps 3 and 6 have overshoot the constraint.

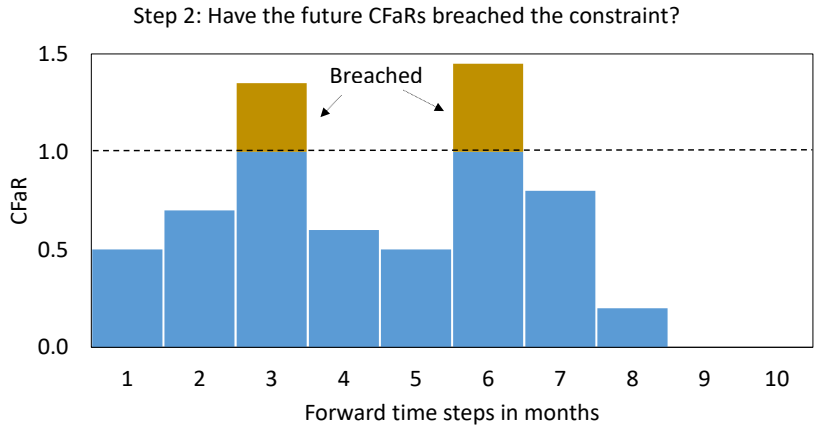

Step 3: Take negative hedges (short domestic currency, long foreign currency) to reduce the breached CFaRs down to the liquidity budget. These negative positions will need to be hedged off by extra positive hedges (long domestic currency, short foreign currency) using other tenors in which the CFaRs have not breached the liquidity constraint.

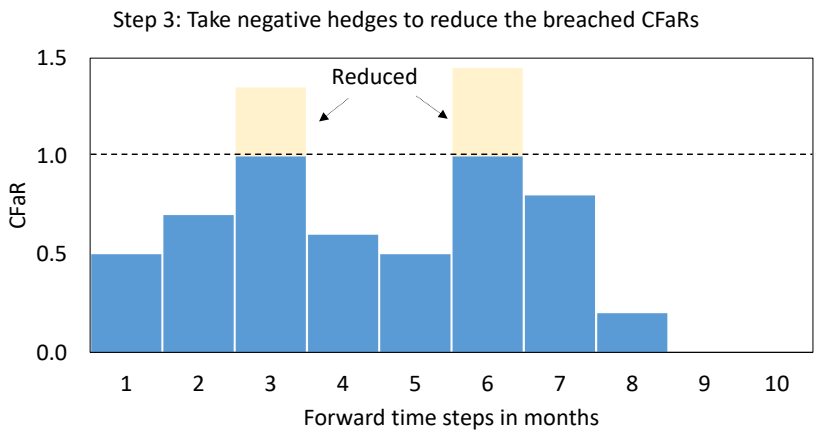

Step 4: (a) Starting from the shortest tenor, time 1, if the CFaR has not been breached, enter the hedge amount such that the newly-estimated CFaR at time 1 matches the liquidity budget $L$; (b) for the second shortest tenor, time 2, if the CFaR has not been breached, enter the hedge amount such that the newly-estimated CFaR at time 2 matches the liquidity budget $L$; (c) repeat this loop forwrads in time until the total nominal hedge amount and the newly added negative positions have been fully hedged. Note that over-hedging is not permitted and the process of adding positions is stopped once the predetermined 100% hedge ratio has been achieved. For example, at time step 10, the newly added CFaR is not raised to match the targeted level, because the nominal amount has been fully hedged before the targeted capacity of CFaR is reached.

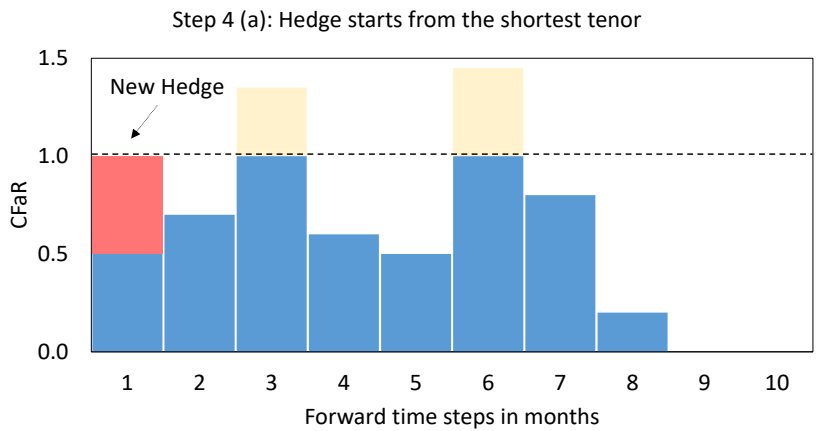

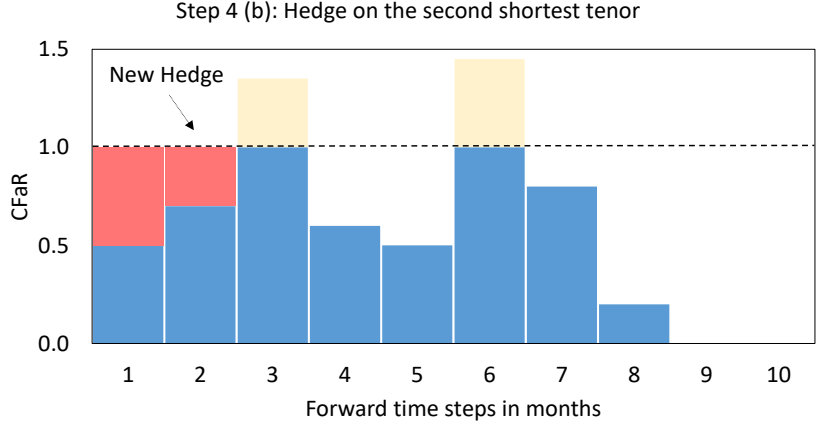

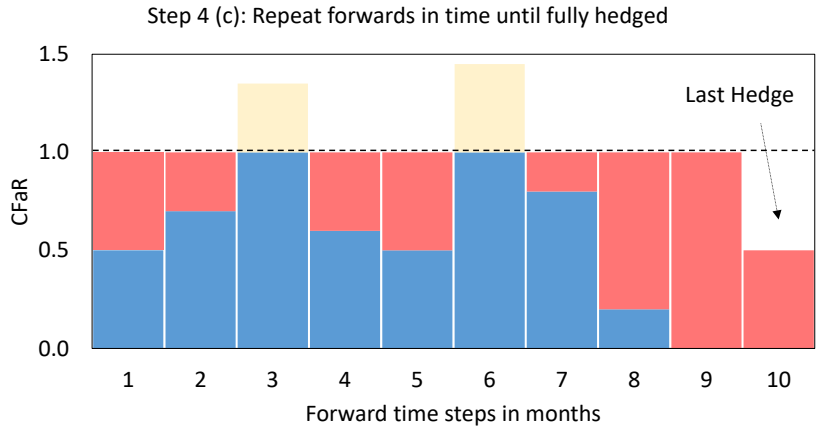

## 7. Additional Constraints

In the hedging strategy, the optimal nominal hedge amount, if the tenor is picked, is determined by

$$a_{t,T} = \frac{L - p\text{-CFaR}_{T|t^-}}{p\text{-CFaR}_{T|t}^{\text{Unit}}}. \tag{20}$$

The issue is that if the current AUD value is far below its long-run, the future $p\text{-CFaR}_{T|t}^{\text{Unit}}$ may become negative. In such a case, the optimal tenor allocation becomes ambiguous.

Another unwanted situation is that when the current AUD value is far above its long-run mean and it tends to depreciate such that a majority of the future CFaRs are breaching the liquidity budget, i.e., $L < p\text{-CFaR}_{T|t^-}$, a large amount of negative hedges will be needed to bring down these CFaRs. On the other side of the same coin, an equivalent extra amount of positive hedges will be needed to maintain the predetermined hedge ratio. In such a case, the nominal hedge values in both short and long positions can be unreasonably high in magnitude.

Here, we provide some corrections for these cases:

$$a_{t,T} = \begin{cases} \min\{a_{t,T}, \bar{a}\}, & \text{if } p\text{-CFaR}_{T|t}^{\text{Unit}} > 0 \text{ and } L \geq p\text{-CFaR}_{T|t^-}, \\ \max\{a_{t,T}, \underline{a}\}, & \text{if } p\text{-CFaR}_{T|t}^{\text{Unit}} > 0 \text{ and } L < p\text{-CFaR}_{T|t^-}, \\ \bar{a}, & \text{if } p\text{-CFaR}_{T|t}^{\text{Unit}} \leq 0 \text{ and } L \geq p\text{-CFaR}_{T|t^-}, \\ 0, & \text{if } p\text{-CFaR}_{T|t}^{\text{Unit}} \leq 0 \text{ and } L < p\text{-CFaR}_{T|t^-}. \end{cases} \tag{21}$$

We regard $\underline{a} = -1$ and $\bar{a} = 1$ as the unconstrained case.

# 8. Analysis on Initial Static Hedge Tenors

The nominal hedge amount depends on various parameters, including the liquidity budget $L$, the CFaRs tail probability $p$, the mean-reversion speed $k$, the mean level $\theta$, the spot volatility $v$, and the current spot level $S_0$. We test the sensitivity of the tenor allocation of the nominal hedge amounts with respect to these parameters. In particular, we use $L=0.01$ in AUD (domestic currency), $p=1\%$, $k=0.4$, $\theta=1/0.75$, $v=0.2$ and $S_0=1/0.75$ as our base parameters, and we vary the value of one of these parameters in the sensitivity analysis, ceteris paribus. We set the current spot price at its mean level to omit the mean-reversion drift that affects the valuation of CFaR. We consider a monthly equally-spaced time grid and we assume that the maximum hedge tenor is 120 month although in the following tests, the hedge tenors will not reach this maximum.

Figure 4 shows that the tenor allocation is overly sensitive with respect to the liquidity budget, although Eq. (20) shows a proportional relationship. This is because the hedging strategy starts filling up the liquidity budget from the shortest tenors, as we can see in the figure that the allocation for short tenors is indeed proportional to the liquidity budget; however, the fully hedged constraint stops the hedging once a 100% hedge ratio has been reached. An investor needs to hedge up to only 4 months if a 5% liquidity budget is used, 16 months if a 0.02 liquidity budget is used, and about 5 years if a 0.01 liquidity budget is used.

**Figure 4** Sensitivity of the tenor allocation with respect to the liquidity budget. Assume $p=1\%$, $k=0.4$, $v=0.2$ and $S_0=1/0.75$. *The lower the liquidity budget, the longer the FX hedges need to be to remain within the liquidity budget.*

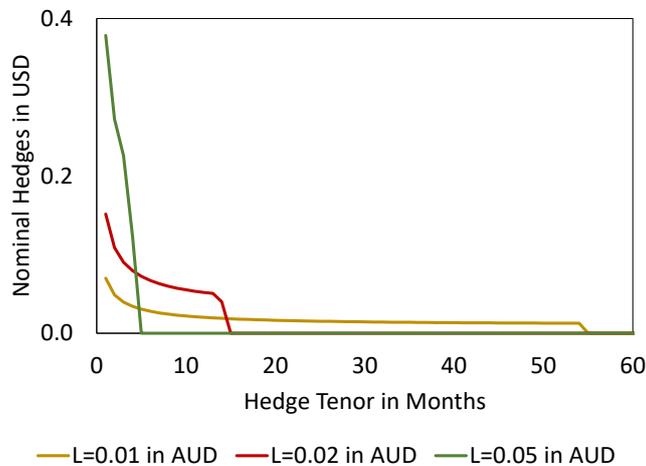

The liquidity risk constraint originally defined in Eq. (2) is given by $p\text{-CFaR}_{T|t} \leq L$. Thus, another key parameter in this constraint, in addition to the liquidity budget, is the tail probability $p$ that defines $p\text{-CFaR}$. Figure 5 shows the sensitivity of the tenor allocation with respect to the value of $p$. An investor needs to hedge up to about 2 years if the tail probability is 5%, 2.5 years if the tail probability is 2%, and about 3 years if the tail probability is 1%. The tenor allocation is not very sensitive to this tail probability. So, to manage the liquidity risk of a foreign investment, the liquidity budget plays a much more critical role than the tail probability.

**Figure 5** Sensitivity of the tenor allocation with respect to the tail probability. Assume $L=0.01$ in AUD, $k=0.4$, $v=0.2$ and $S_0=1/0.75$. *The lower the tolerance for breaching the CFaR limit, the longer the hedges need to be to remain within the liquidity budget.*

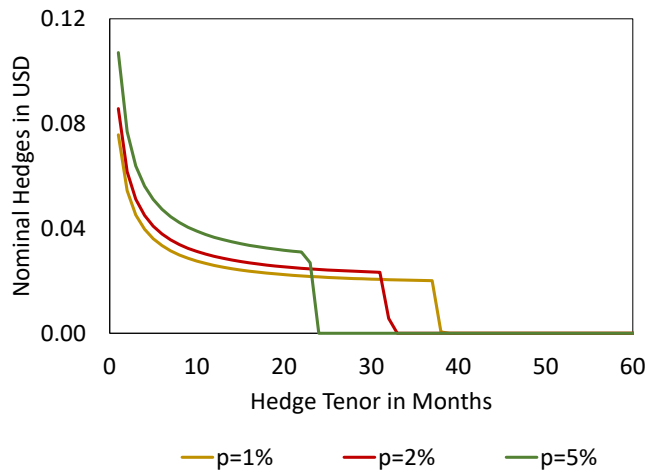

We start our sensitivity analysis by first looking at the spot volatility. Figure 6 shows that the tenor allocation is very sensitive to the spot volatility. The higher the spot volatility, the higher the CFaR for hedging one unit of USD, and therefore the smaller the allocation of the nominal hedge amount to each tenor. An investor needs to hedge up to about 1 year if the spot volatility is 0.1, about 3 years if the spot volatility is 0.2, and more than 5 years if the spot volatility is 0.3.

**Figure 6** Sensitivity of the tenor allocation with respect to the spot volatility. Assume $L=0.01$ in AUD, $p=1\%$, $k=0.4$, and $S_0=1/0.75$. *The higher the volatility of spot FX, the longer the hedges need to be to remain within the liquidity budget.*

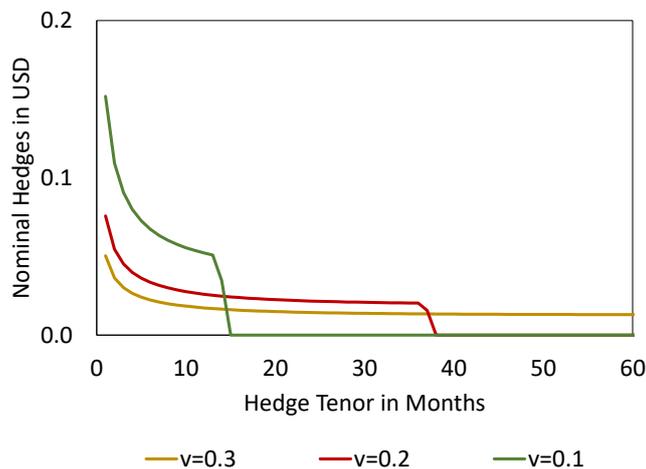

Figure 7 shows the sensitivity of the tenor allocation with respect to the spot price, quoted in USD per AUD. When the spot price is low (AUD is strong), the spot price tends to increase (AUD depreciates), and thus the expected payoff $F_{0,T} - E_0(S_T)$ decreases, which leads to a larger $p$-$\text{CFaR}_{T|0}$. In this situation, only smaller nominal values can be allocated to each tenors, and therefore longer tenors are needed to share the spread-out of the total hedge amount. As shown in the figure, an investor needs to hedge up to about 1.5 years if the spot price is 1/0.50 (AUD is weak), about 3 years if the spot price is 1/0.75 (at the mean level), and about 5 years if the spot price is 1/1.00 (AUD is strong). When the spot price is 1/0.50, it may not revert to the mean level in a short period of time, but it must in the long run. This is the reason why the nominal hedge allocation decreases with tenor in the short run, while increases in the long run.

**Figure 7** Sensitivity of the tenor allocation with respect to the spot price. Assume $L = 0.01$ in AUD, $p = 1\%$, $k = 0.4$, and $v = 0.2$. *The higher the FX rate, the longer the hedges need to be to remain within the liquidity budget.*

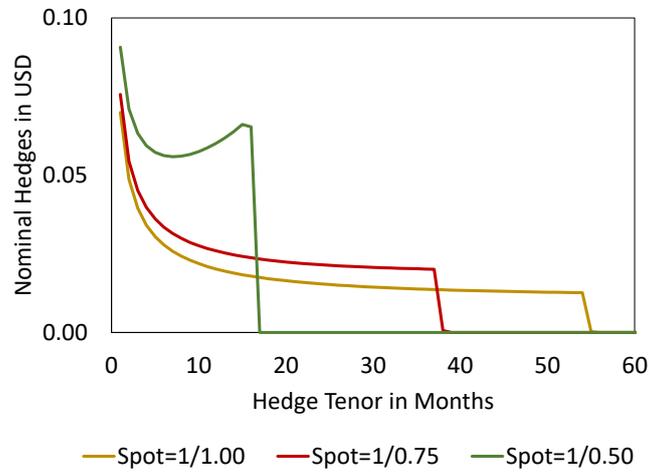

**Figure 8** Sensitivity of the tenor allocation with respect to the mean-reversion speed, with different spot levels. Assume $L = 0.01$ in AUD, $p = 1\%$, and $v = 0.2$.

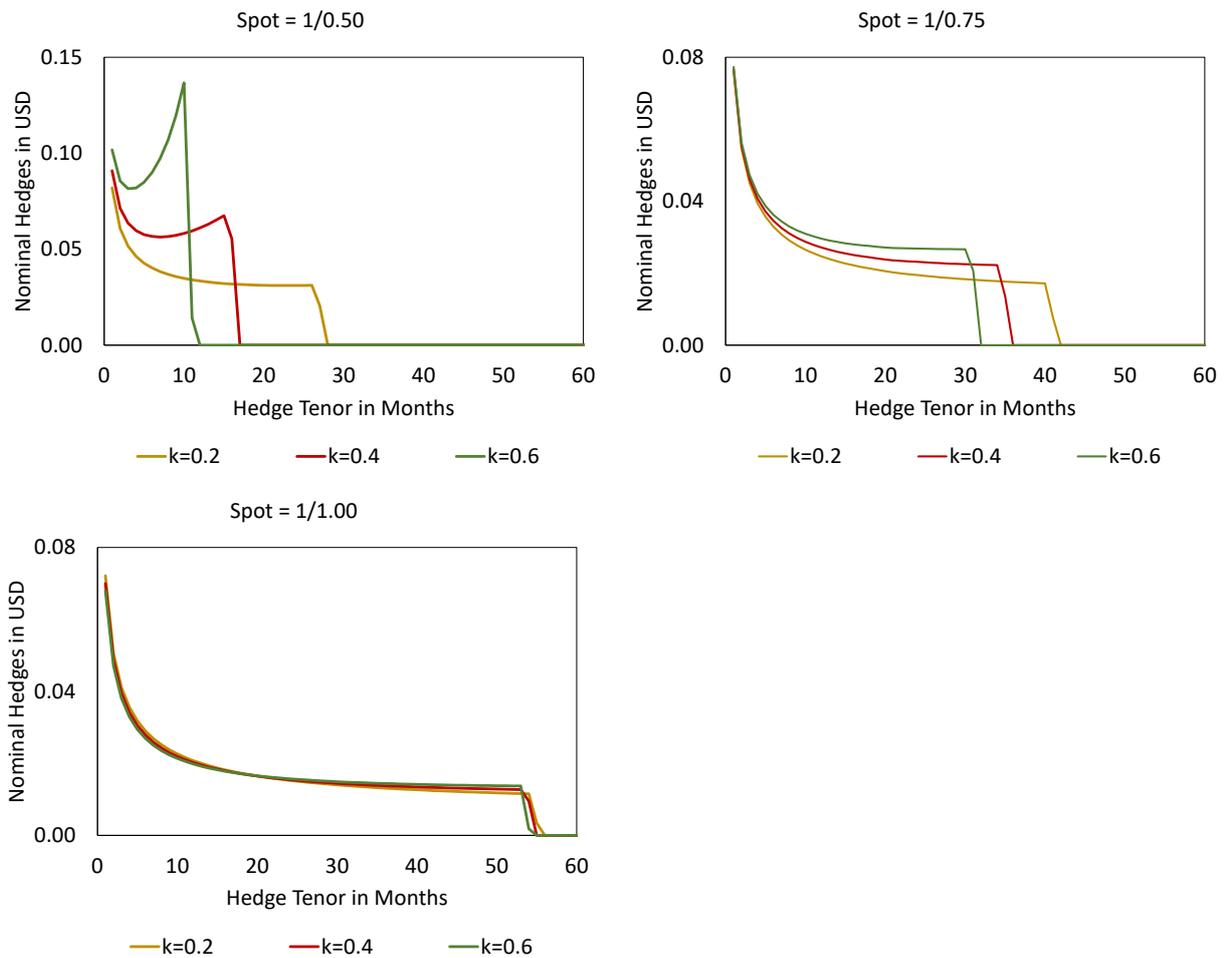

Figure 8 shows the sensitivity of the tenor allocation with respect to the mean-reversion speed, with three different spot levels $S_0 = 1/0.50$ (AUD is weak), $S_0 = 1/0.75$ (AUD is at its long-run mean), and $S_0 = 1/1.00$ (AUD is strong). The results show that the sensitivity effects gradually diminish when AUD appreciates. When AUD is weak, the tenor allocation is relatively sensitive to the mean-reversion speed: the quicker the spot price reverts to the mean level, the lower the CFaR for a short tenor. An investor needs to hedge up to about 2 years if the mean-reversion speed is 0.2, about 1.5 years if the mean-reversion speed is 0.4, and about 1 year if the mean-reversion speed is 0.6. When AUD is strong, the spot price tends to increase (AUD depreciates) and the hedging strategy tends to have negative cash flows, which increases the CFaR. In this case, the tenor allocation becomes invariant to the mean-reversion speed, indicating that when the hedging strategy tends to be out-of-money, the mean-reversion speed does not affect the overall liquidity risk as long as the spot price does tend to revert to the mean level.

## 9. Dynamic Hedge Tenor

The allocation of hedge tenors in a dynamic context is more complicated than that in a static context. In a static context, the CFaRs in all the future time buckets are null, and therefore it is simple. In a dynamic context, the CFaRs are non-zero, and some may breach the liquidity budget. In the implementation, we need to keep track of the forward rates and the tenors of all the forward contracts until expiry.

We provide simulation study for the dynamic hedging. We consider a 20-year horizon and a monthly-spaced time grid. We use 10,000 simulation paths. We assume mean-reversion speed $k = 0.4$, mean-reversion level $\theta = 1/0.75$, volatility $v = 0.2$ and initial spot price $S_0 = 1/0.75$. We use $p = 1\%$ for tail probability of CFaR and $L = 0.01$ AUD for liquidity budget. We also assume that the shorter-dated forwards generate higher carry returns and we fill up the liquidity budget starting from the shortest tenor, as described in Section 5. Regarding the forward rates in the simulation, we obtain the historic forward rates from November 1993 to August 2018, and we use the average of spot-to-forward ratio as a fixed ratio to generate forward rates given simulated spot prices. For the tenors whose forward rates are not directly observed on the forward curve, we estimate a proxy by using a linear interpolation between its nearest two tenors. Note that we assume a fixed spot-to-forward ratio only for the demonstration of the simulation study. The optimal tenor derived from the analytic model is in closed-form expression which does not rely on Monte Carlo simulation, thus there will not be any assumption on the forward curve when this optimal strategy is implemented in practice.

**Figure 9** Dynamic Hedging – AUD Cash Flow Distribution

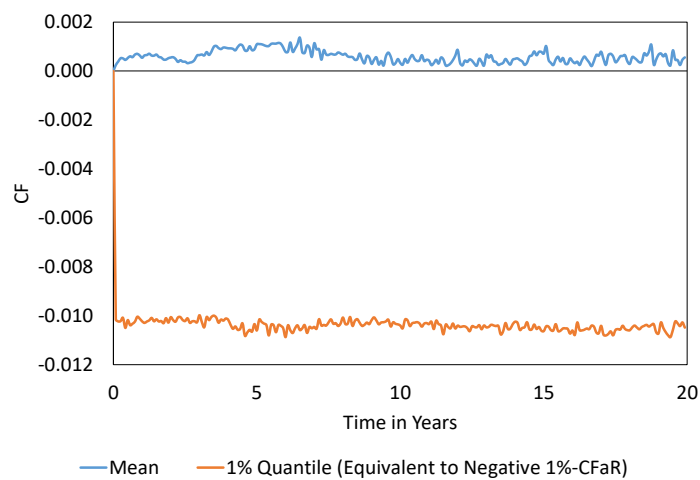

**Figure 10** Dynamic Hedging Expected Nominal Hedges in USD

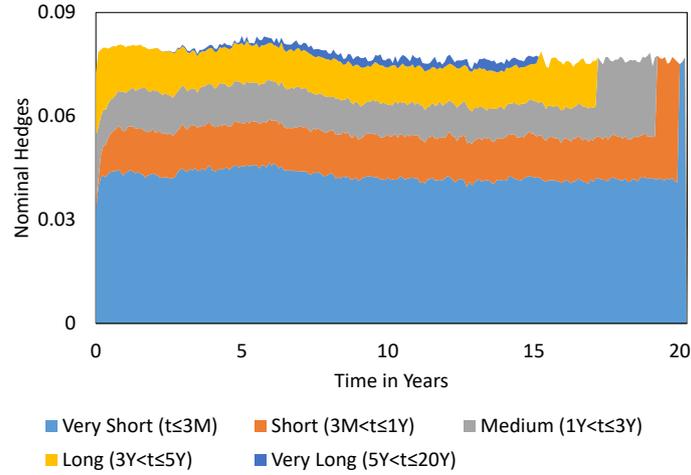

Figure 9 shows the expected value and the 1% quantile of monthly cash flows over the 20-year time. The 1% quantile of cash flow is equivalent to the negative 1%-CFaR. The objective is to maintain this quantity within the liquidity budget $L=0.01$ AUD. As shown in the figure, the CFaR is well managed at about 0.01, which indicates the algorithm's capability to achieve what is required by the liquidity constraint. Figure 10 shows the expected tenor allocation of the nominal hedges. A majority of nominal hedges is allocated to very short tenors. This is because the CFaRs are small in short periods, for which large nominal hedges can be used to generate high carry returns.

## 10. Practice and Backtesting

In the previous sections, we assume for demonstrative purpose only, that shorter-dated tenors tend to deliver higher carry trader returns. In this section, we remove this assumption. Instead, we sort the tenors according to their expected carry trade return which consists of forward points, transaction costs as well as the expected change in the spot price. We backtest our optimal hedging strategy, and benchmark it with an equal weight strategy.

### Data and Calibration

We obtain monthly close spot prices of AUD per USD, spanning from November 1993 to August 2018. In order to have explicit expressions of expected carry trade and CFaR, we need to model USD per AUD. Thus, we calibrate the Ornstein Uhlenbeck process in Eq. (4) to the inverse spot prices, and we obtain the estimated mean-reversion level $\hat{\theta}=1/0.7549$, mean-reversion speed $\hat{k}=0.2139$ and volatility $\bar{v}=0.1627$. We also obtain monthly close forward curves, for which we use linear interpolations to estimate the forward rates for the missing tenors. All these data are obtained from Bloomberg.

We consider in-sample backtesting. Our objective is to investigate how the hedging strategy works assuming that the FX rate is mean-reverting. An out-of-sample backtesting will be more realistic and more useful for practitioners. However, in-sample backtesting sheds more light on the hedging strategy risk metric and provides more intuitive insights on why the strategy works, without the additional random errors from the out-of-sample predictions.

## Transaction Costs

We also consider transaction costs in the backtesting study. The transaction costs for different tenors are given in Table 1. We use linear interpolation between every two tenor to extract the transaction costs for each monthly increment.

**Table 1** Annualized Proportional Transaction Costs

| 3M | 1Y | 2Y | 3Y | 5Y | 7Y |
|---|---|---|---|---|---|
| -0.01% | -0.02% | -0.04% | -0.05% | -0.08% | -0.10% |

## Expected Carry

The systematic hedging strategy demonstrated in previous sections is built upon an assumption that rolling shorter tenor FX forwards delivers higher carry trade returns than longer tenor FX forwards. Although this assumption is valid on average in this dataset, it is not always true in general. For example, regardless of the carry from the spot price, the carry returns on the forward points do not decrease with tenor in periods in late 1990s and in the recent years 2018-2019, see Figure 11 for example.

**Figure 11** Annualized Carry on Forward Points

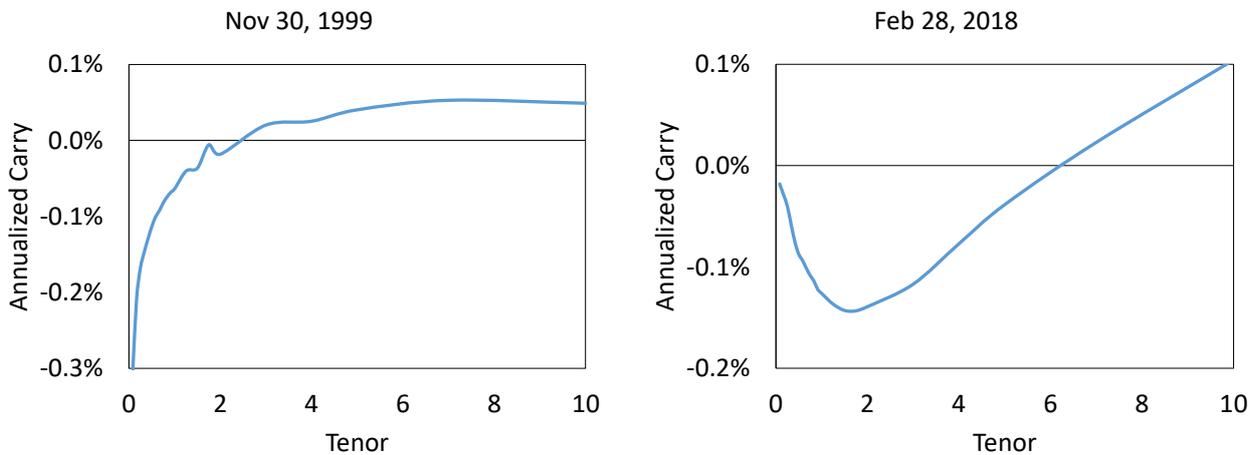

Moreover, due to the mean-reverting property of the spot price, the expected carry trade return caused by the spot price movement is also time-varying. When AUD drops below its long-run mean, it tends to appreciate, and therefore the expected carry trade return from the spot component will increase and vice versa. Here, we use a systematic ranking strategy to rank the tenors by their expected carry returns, taking into account of both the carry trade returns observed on the forward curve and the expected spot price according to it mean-reverting property. To perform the hedging strategy, we first rank the tenors by their expected carry trade returns subtracted by transaction costs, $(F_{t,T} - E[S_T])/(T-t) - \text{TC}_{t,T}$, and then we follow the algorithm in Section 5 to spread out the nominal hedge amounts into different tenors according to the ranking.

## Example

We consider a strategy without position constraints, i.e., $\underline{a} = -1$ and $\bar{a} = 1$. We use $L = 0.01$ AUD for the liquidity budget. We show how our hedging strategy is able to generate cash inflows once AUD appreciates and how it is able to maintain the cash outflows within the predetermined liquidity budget.

We first look at the historic spot prices in Figure 12. The spot price experiences its historic low at about 0.5 USD per AUD during 2000, at which the spot price is expected to mean-revert to at least the historical long-run equilibrium at about 0.75. The spot price does increase in the following years, and cash inflows can be expected until the spot price stops its upward momentum in 2003. Therefore, large nominal hedge amounts are entered to long AUD during the period from 2000 to 2003 (see Figure 13), and consequently, large cash inflows are collected during this period (see Figure 14).

During 2008 and 2009, AUD depreciated more than 30%, before rebounding strongly, as shown in Figure 12. A very possible but highly undesirable situation for every fund manager is that they are required to sell their assets to fund the hedge settlements once the domestic currency has fallen significantly, forcing them to suspend currency hedging in fear of further falls. Like the MTAA case discussed in the Introduction, this can be catastrophic if the domestic currency then rebounds causing massive losses for their now-unhedged portfolio should the domestic currency rebound. Our hedging strategy is able to minimise such a risk to a specified degree of statistical confidence. By spreading out the liquidity risk and the nominal hedge value widely into different time buckets, our hedging strategy only needs to settle small cash outflows in any short period of time, and Figure 13 shows that the liquidity risk can be well managed as required.

After AUD hits its historic high at about 1.1 in 2011, it continuously drops in value in the following years from 2011 to 2014. Such drop will generate substantial short-term cash outflows if short-dated hedge tenors are used. However, the hedging strategy tends to use long-dated tenors to avoid the potential of large short-term cash outflows. As a consequence, since the long-dated FX forwards contracts are not expiring in the following years, only very small nominal hedge amounts are entered to maintain a 100% hedge ratio (see Figure 13) and therefore the cash outflows can be largely deferred and even partly eliminated when the trajectory of AUD reverses (see Figure 14).

Figure 15 shows the cumulative P&L of the offshore investment portfolio in AUD, where for simplicity we assume zero discount rate. Because we assume the offshore asset value $W=1$ USD constant over time, the P&L of the unhedged portfolio value is proportional to USD per AUD. Regarding the P&L of the hedged portfolio, we add cash flows from the currency hedging to the P&L of the unhedged portfolio. The result shows that, taking into account of the aforementioned advantages, over a long time, the hedged portfolio significantly outperforms the unhedged portfolio. By taking into account of the mark-to-market (MtM) value of the not-yet-expired hedges, the hedged P&L shows a steadily increasing trend with little fluctuation over time.

**Figure 12** Spot Price - AUD per USD

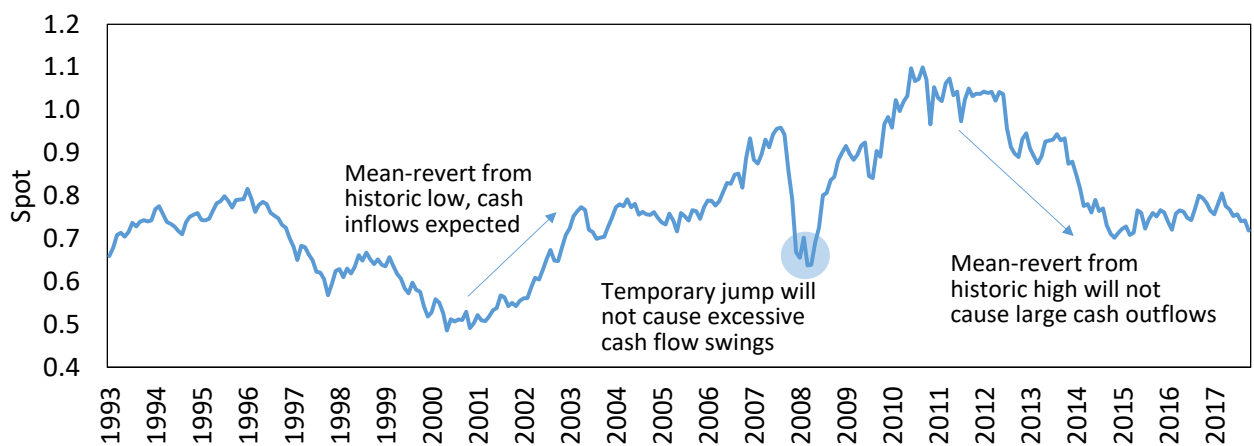

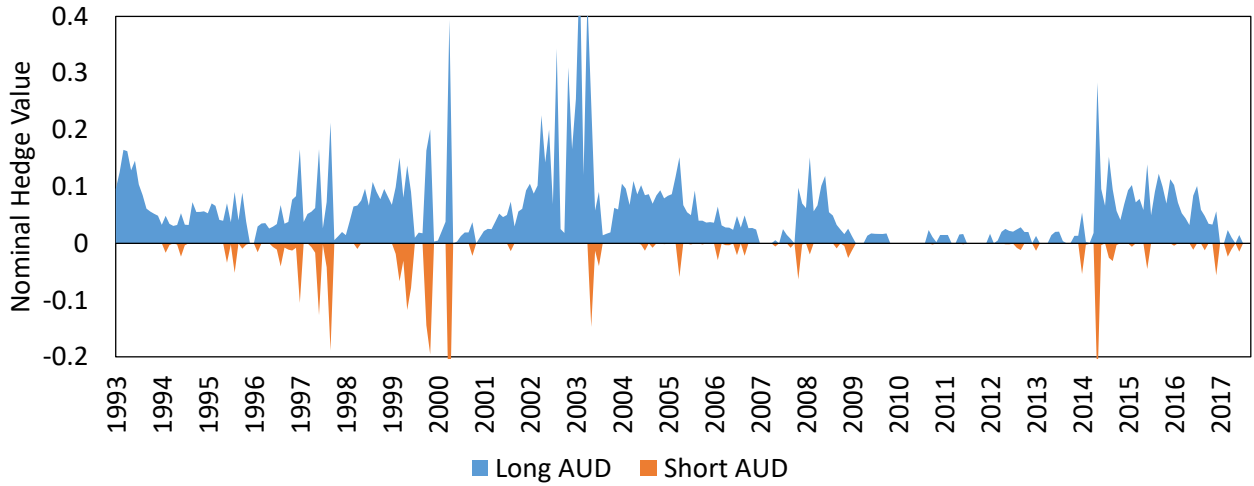

**Figure 13** Total Long/Short Nominal Hedge Values in USD

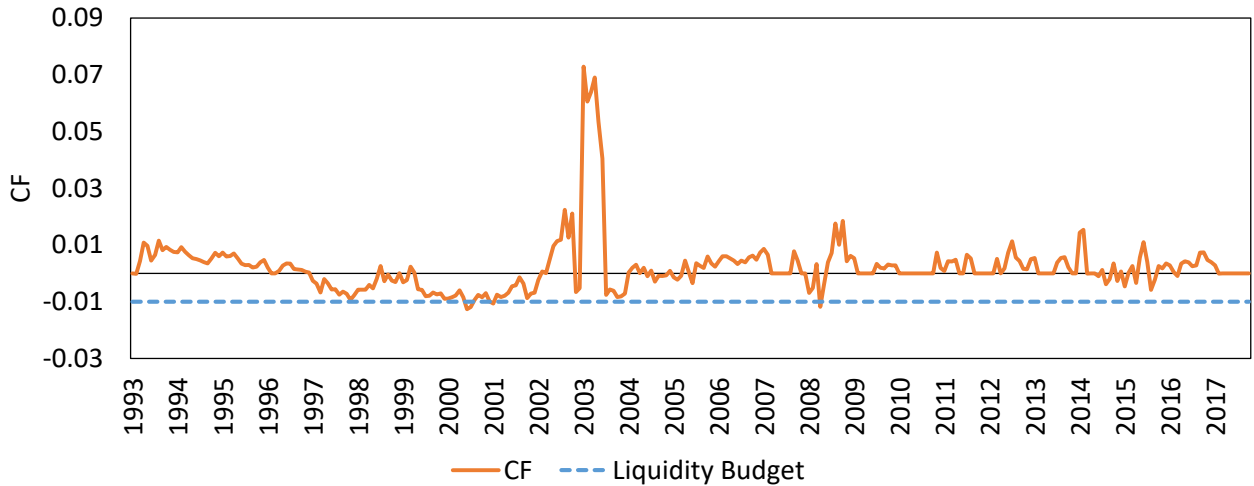

**Figure 14** Monthly AUD Cash Flows from Hedging

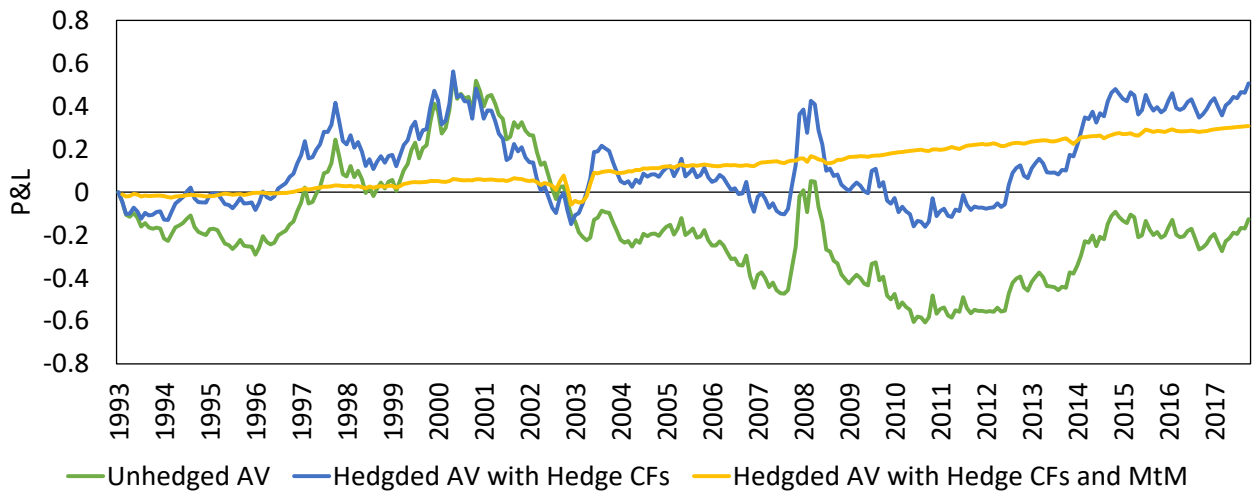

**Figure 15** Cumulative P&L of Investment Portfolio Valued in AUD

**Benchmark**

We compare our hedging strategy to several benchmark equal weight strategies. The equal weight strategy is as follows: suppose we wish to spread out the nominal value into monthly time buckets up to one year, the equal weight strategy assigns 1/12 nominal hedges equally spread out into each month; once a nominal hedge of 1/12 expires, the strategy enters another one-year forward contract with nominal value of 1/12; and as time rolls, the total nominal value is always spread out equally into every month in the following one year time. If the hedge tenor is two year, the equal weight strategy spreads out the hedge into 24 nominal values of 1/24.

We consider the following settings in our hedging strategy:
1. Unconstrained long/short strategy with $\underline{a}=-1$ and $\bar{a}=1$, and the liquidity budget is $L=0.01$ AUD,
2. Constrained long/short strategy with $\underline{a}=-0.01$ and $\bar{a}=0.1$, and the liquidity budget is $L=0.01$ AUD,
3. Unconstrained long-only strategy with $\underline{a}=0$ and $\bar{a}=1$, and the liquidity budget is $L=0.02$ AUD,
4. Unconstrained long-only strategy with $\underline{a}=0$ and $\bar{a}=1$, and the liquidity budget is $L=0.01$ AUD,
5. Unconstrained long-only strategy with $\underline{a}=0$ and $\bar{a}=1$, and the liquidity budget is $L=0.005$ AUD,
6. Unconstrained long-only strategy with $\underline{a}=0$ and $\bar{a}=1$, and the liquidity budget is $L=0.002$ AUD.

Table 2 summarizes the results of the backtested cash flows from different strategies. Given a similar level of carry trade return, the proposed hedging strategy delivers a smaller 1%-CFaR, compared to the equal weight strategy. Symmetrically, given a similar level of 1%-CFaR, the proposed strategy delivers a high carry trade return. As expected, the unconstrained long/short strategy (Str1) delivers the highest carry trade returns, and the constrained long/short strategy (Str2) delivers the second highest. For the long-only strategies, the smaller the liquidity budget, the smaller the 1%-CFaR and the smaller the carry trade return.

**Table 2** Backtesting Results of Monthly Cash Flows in AUD, per 100 units of USD

|  | Test Strategies | | | | | | Benchmark Strategies | | |
|---|---|---|---|---|---|---|---|---|---|
|  | Str1 | Str2 | Str3 | Str4 | Str5 | Str6 | Eq 1Y | Eq 3Y | Eq 10Y |
| An. CF | 2.55 | 2.49 | 2.33 | 2.24 | 2.18 | 2.02 | 2.33 | 2.05 | 1.88 |
| Volatility | 3.50 | 3.45 | 3.82 | 2.28 | 1.97 | 1.14 | 5.04 | 2.75 | 0.97 |
| 1%-CFaR | 1.08 | 0.99 | 2.23 | 1.14 | 0.58 | 0.24 | 3.43 | 1.33 | 0.36 |
| min | -1.26 | -1.25 | -3.59 | -1.59 | -0.77 | -0.40 | -3.69 | -1.55 | -0.42 |
| max | 7.29 | 8.14 | 5.79 | 3.37 | 3.95 | 3.66 | 3.90 | 2.10 | 0.91 |

## 11. Conclusion

This paper provides an optimal FX forward hedging framework, assuming mean-reversion of the FX spot rate and a 100% FX hedge ratio. The framework determines a set of hedge tenors that optimally balances FX carry returns and liquidity risk. The liquidity risk is measured by the Cash Flow at Risk (CFaR) metric which is simply the Value at Risk (VaR) of the cash flow generated from the hedging. Given a liquidity constraint that the CFaRs in all the future time buckets should not breach a predetermined liquidity budget, the strategy spreads out the nominal hedges into the tenors that have the highest (lowest) expected carry benefit (cost). In our examples we have assumed that AUD is the fund manager's domestic currency and USD is the foreign currency.

We demonstrate the insights by first considering the case where shorter FX forward tenors have the higher expected carry returns, which is the fact on average in the history of USD/AUD currency pair. The optimal hedge tenor is very sensitive to the liquidity budget: the lower the liquidity budget, the longer the tenors can be. Similarly for sensitivity to the FX spot rate volatility: the higher the volatility, the longer the tenor must be. The FX spot rate, its mean-reverting speed and its mean-reverting level also jointly affect the choice of tenors.

We show by Monte Carlo simulation and by in-sample backtesting that our hedging strategy does indeed protect the cash flows within the liquidity budget. In the backtesting, we introduce a tenor-ranking method which is useful when the "short-first" assumption fails. We show that the proposed hedging strategy can accumulate cash inflows when AUD appreciates; avoid incurring large cash flows over a short time jump-and-rebound; and avoid large cash outflows during a prolonged depreciation.